\documentclass[a4paper]{article}

\usepackage{INTERSPEECH2020}
\usepackage[linesnumbered,ruled,vlined]{algorithm2e}
\usepackage{algorithmic}
\usepackage{hyperref}

\usepackage{defs,semicruch}
\SetArgSty{textnormal}
\SetAlCapNameFnt{\small}
\SetAlCapFnt{\small}
\SetAlgorithmName{Algo}{Algo}{List of Algorithms}
\newcommand{\algname}{FineMerge}
\newcommand{\vq}{\vek{q}}

\title{Black-box Adaptation of ASR for Accented Speech}
\name{Kartik Khandelwal, Preethi Jyothi, Abhijeet Awasthi, Sunita Sarawagi}
\address{
  Indian Institute of Technology Bombay, Mumbai, India}
\email{\{kartikk,pjyothi,awasthi,sunita\}@cse.iitb.ac.in}

\begin{document}

\maketitle
\begin{abstract}
We introduce the problem of adapting a black-box, cloud-based ASR system to speech from a target accent. While leading online ASR services obtain impressive performance on mainstream accents, they perform poorly on sub-populations --- we observed that the word error rate (WER) achieved by Google's ASR API on Indian accents is almost twice the WER on US accents. Existing adaptation methods either require access to model parameters or overlay an error correcting module on output transcripts.  We highlight the need for correlating outputs with the original speech to fix accent errors.  Accordingly, we propose a novel coupling of an open-source accent-tuned local model with the black-box service where the output from the service guides frame-level inference in the local model. Our fine-grained merging algorithm  is better at fixing accent errors than existing word-level  combination strategies.  Experiments on Indian and Australian accents with three leading ASR models as service, show that we achieve upto 28\% relative reduction in WER over both the local and service models.
\end{abstract}

\noindent\textbf{Index Terms}: Black box ASR systems, accented speech recognition, adaptation.

\section{Introduction}

The emergence of cloud-based AI services, for tasks like machine translation and speech recognition, have greatly increased the accessibility of machine learning. These services are powered by sophisticated engines and trained on large proprietary datasets. The internals of these engines are often not exposed to clients. Often   
a client's input comes from a different domain than the training domain of the server.  The existing fix of retraining to adapt to new target domains is not an option in this case.
This leads us to our problem of \emph{Black-box Adaptation.}

In this work, the task of interest is automatic speech recognition (ASR) in English, and the domains correspond to different accents. While leading online services like the Google ASR API~\cite{GoogleASR} attain superior performance
on high-resource English accents, they perform poorly on a large number of under-represented English accents. The API gives word error rates (WERs) of $23\%$ or higher on datasets in Australian and Indian accents, as opposed to a WER of
$13.2\%$ on US accents. 
As ASR systems start getting deployed in several critical applications, it is increasingly imperative to design light-weight methods of accent adaptation to provide fair access to users of all regions and ethnicity~\cite{Koenecke7684}.
%
Existing methods of adapting ASR models to a specific accent~\cite{huang2014multi,chen2015improving,yang2018joint,jain2018improved,viglino2019end,rao2017multi},  require modifying
model parameters, which is not an option in the black-box setting. 

One could implement black-box adaptation in the form of an error-correction model to alter the outputs from the service~\cite{ringger1996error}. When the mismatch is in the language model, error correction models using a domain-specific language model have been proposed before~\cite{corona2017improving}.   
However, for
recovering from accent errors an error correction model would need to correlate the service's transcript with the original speech.
To handle speech, the model would in turn need to incorporate an ASR system.  This leads us to switching our perspective, so that black-box adaptation amounts to
building a local ASR system which is retargeted to correct accent errors in the service's output. 


The local ASR system would be an open-sourced ASR architecture like
DeepSpeech2~\cite{DS2} pretrained on a publicly available corpus  like the US-accented Librispeech~\cite{Librispeech} corpus
but further finetuned using a small amount of data in the target accent. 
Typically the local model would be less accurate than the service in all parts except the parts with systematic accent differences. 
If the outputs from the local and service models are combined via standard  combination approaches at the word or transcript-level ~\cite{rover,soto2016selection}, we obtain only limited improvements in accuracy over the service. In other words, if the local system were also to be used as a black-box, we would not obtain the performance improvements we seek.

Hence, we exploit our white-box access to the local system.  Our idea, at a high-level, is to use the transcript obtained from the service to \emph{guide the inference} of the local ASR system.  Our guided inference algorithm (named \algname)
aligns the characters in the service with input frames using a Viterbi-like decoding and then selectively modifies the frame-level distribution of the local model.  Our fine-grained merging step is easy to plug in existing speech pipelines, fast during inference, and 
specifically tailored to fixing accent errors --- we often recover words that were absent from stand-alone outputs of both local and service models.
Experiments on different service APIs on two different English accents show that \algname\ provides significant reduction in WER over either the local or service models, and existing methods of combining them at the word-level.  To summarize, our overall contribution in this paper are:
\begin{enumerate}
    \item We introduce the problem of black-box accent adaptation of ASR service APIs.
    \item We propose an efficient coupling of a local white-box model with a black-box service to accent adapt with limited labeled data without incurring the cost of accessing the service during training.
    \item We design a novel guided inference algorithm on the local model that is specifically tailored to correct focused accent errors in an otherwise strong service API. 
    \item We evaluate our algorithm on two accents and three service APIs.  Our approach provides up to 28\% reduction in relative WER over both local and service models. Existing methods based on rescoring N-best lists or combining outputs at the word-level are not as effective. 
\end{enumerate}

\section{Related Work}
\noindent{\bf Accent Adaptation in Speech.}
Accent adaptation in speech 
has been a problem of long standing interest. 
%
One category of methods attempt to create accent invariant systems 
and range from early approaches that simply merged data from multiple accents for training a single model~\cite{vergyri2010automatic} to more recent work that uses adversarial learning objectives to extract accent-invariant feature representations from speech~\cite{sun2018domain,chen2019aipnet}. 
A second category of methods are accent dependent methods that adapt to the speaker's accent.   
Early approaches were HMM-based acoustic model adaptation and pronunciation model augmentation with accent-specific pronunciations~\cite{humphries1996using,zheng2005accent}. Within neural models, accent adaptation was achieved via accent-specific output layers~\cite{huang2014multi,chen2015improving} and hierarchical models in a multitask learning setting~\cite{rao2017multi}. A more recent work jointly learns an accent classifier and accent-dependent models~\cite{yang2018joint,jain2018improved,viglino2019end}. Our method is also accent dependent but we need to adapt a black-box service model. We build local accent-adapted ASR systems, which are in turn guided during inference by service predictions.

\noindent{\bf Black box ASR Systems.} Speech transcription services have seen widespread use in recent years. However, the underlying ASR systems in these services are black box systems. Adapting such models to a client's needs would be of great utility but prior work in this area is sparse. \cite{watanabe2014black} shows how to optimize black box ASR systems. 
and \cite{kastanos2019confidence} shows how to 
improve confidence estimates produced by such black-box systems.  Another closely related work~\cite{corona2017improving} is to use a domain-specific language model and a semantic parser to rescore the hypotheses from a black-box ASR system. Unlike their method, we 
achieve a more fine-grained integration of our client model with the service.

\noindent{\bf System Combination Approaches.} Ours can be viewed as a type of system combination approach which has seen wide use in ASR. ROVER (Recognizer Output Voting Error Reduction)~\cite{rover} is one of the most popular techniques that first combines predictions from different systems using an alignment step followed by a weighted voting step. Prior work on dialectal speech recognition~\cite{soto2016selection} observed that using the best output from a dialect-specific model is more accurate than techniques like ROVER. 
Unlike ROVER that considers each individual system as a black-box, our method that leverages white-box access to a local accent-adapted ASR system is more targeted to correct accent errors and ultimately more accurate.

\section{Our Approach}
Given an audio signal $\vx$, we invoke the service model $\cS$ on $\vx$ and get the transcript $\vs$ comprising of tokens $s_1,\ldots,s_k$, along with token-level confidences $p_1,\ldots,p_k$.  In addition the client can invoke a  local white-box model $\cC$ that has been trained/fine-tuned on a limited accented labeled data.  On input $\vx$, let $\vc = c_1,\ldots,c_r$ denote the transcript from the local model $\cC$ with 
token-level confidences $\vq=q_1,\ldots,q_r$.  In general the number of tokens in the two outputs ($k,r$) could be different.

One option to merge the transcripts of the two models is using a word-level aligner like Rover~\cite{rover}.  However, for accent errors we expect the service to be wrong only on a sub-part of a word, say a 't' being wrongly identified as a 'd'.  The local transcript $\vc$ might correct some accent errors while missing out on other parts of the word.  In general, the local model is expected to be weaker than the service on all but the accent errors, for the client to want to pay for the service.  As an example consider the first sentence in Table~\ref{tab:anec} showing the gold transcript $\vy$, service transcript $\vs$, and local transcript $\vc$ for an Indian accented model.  The service model fails to recognize the {\tt t} in {\tt toasted} and outputs {\tt posted}.  The local model recognizes the {\tt t} but yields {\tt to\,state}.  To reconstruct the correct word in such cases we need a finer-grained splicing at sub-word level.

Given the prevalence of character-level models in modern ASR systems we then sought to splice the two transcripts at the character-level.
Designing a good character-level merging strategy is challenging because of large divergences between the two outputs both because of the differential strengths of their acoustic models and the introduction of unheard characters when biasing with their respective language models. Strategies like combining the characters from the two outputs using Rover-like algorithms fail to distinguish between the two types of errors in the absence of accurate character-level confidence from the service.
%
For example, aligning the characters in {\tt posted} with {\tt to\,state} yielded {\tt too\,sttd} 

We finally designed an algorithm that exploits white-box access to the local model $\cC$ to guide its decoding using the service transcript $\vs$, instead of merging a fixed $\vc$ from $\cC$.

We assume the local model $\cC$ is trained using the standard 
CTC loss invoked on frame-level character distributions~\cite{graves2006connectionist} that maximizes likelihood of the target $\vy$ by marginalizing over all alignments compatible with $\vy$.
During inference, the trained model generates the distribution over alignments for an input $\vx$ and predicts character distributions $P_1,\ldots,P_T \triangleq \vP$ at each of the $T$ frames of the input. From these probability distributions, an output sequence $\vc$ is recovered using beam-decoding in conjunction with a language model (LM). 



We guide this inference using the service transcript in two steps:   First align the service characters with each frame of the local model using its frame-level probability distributions $\vP$.  Next revise $\vP$ to selectively support $\vs$.  We elaborate these steps next.  A pseudo code appears in Algorithm~\ref{algo:mix_prob}.

\begin{table}[]
\setlength\tabcolsep{2.0pt}
    \centering
    \begin{tabular}{|c|l|c|c|c|c|c|c|c|c|c|c|} \hline
        1 & $S_t$ & \_ & p & o & \_ & \_ & s & t & e & d & d \\ 
        & $P_t(S_t)$ & 6e-5 & 1e-11 & 1 & 0.34 & 0.01 & 0.93 & 0.99 & 0.44 & 0.29 & 0.98 \\ \hline
        2 & $d_t$ &t & t & o & o & & s & t & a & t & d \\ 
        & $P_t(d_t)$ & 0.99 & 0.99 & 1.0 & 0.63 & 0.98 & 0.93 & 0.99 & 0.55 & 0.64 & 0.98 \\ \hline
        3 & $r_t$ & t & t & o & \_ &  & s & t & e & d & d \\ 
        & $P^s_t(r_t)$ & 0.62 & 0.99 & 1.0 & 0.59 & 0.61 & 0.93 & 0.99 & 0.66 & 0.57 & 0.98 \\
        & $P_t(r_t)$ & 0.99  & 0.99 & 1.0 & 0.34 & 0.98 & 0.93 & 0.99 & 0.44 & 0.29 & 0.98 \\ \hline

        & Frame $t$ & 1 & 2 & 3 & 4 & 5 & 6& 7& 8& 9& 10 \\ \hline
    \end{tabular}
    \caption{Example: Client model revising frame-level character distribution $\vP \rightarrow \vP^s$ using service transcript $\vs$='posted' in \algname.   $d_t=\displaystyle\argmax_c P_t(c)$ and $r_t=\displaystyle\argmax_c P^s_t(c)$. First row shows aligned service characters and their probability from $\vP$, second row shows the modes of the $\vP$ distribution, third row shows the argmax $r_t$ of the revised distribution and its probability from the revised and original distribution. }
    \label{tab:example}
\end{table}
\setlength\tabcolsep{5.0pt}
\newlength\mylen
\newcommand\myinput[1]{%
  \settowidth\mylen{\KwIn{}}%
  \setlength\hangindent{\mylen}%
  \hspace*{\mylen}#1\\}

\begin{algorithm}
\DontPrintSemicolon 
\KwIn{$\vx$: Input audio with $T$ frames \\
\myinput{$\cC$: Local model fine-tuned on target accent} \\
\myinput{$\psi$: service probability threshold}\\
\myinput{$\omega$: service weight for mixing}\\
\myinput{$\gamma$: probability of blank}}
\KwOut{Final transcript}
$\vs,\vp \gets $  Transcript, token-confidence from Service on $\vx$\;
$P_1,\ldots,P_T\gets$ Frame-level probability from $\cC(\vx)$\;
$S_1,\ldots,S_T \gets$ Viterbi-align($\vs, \text{Smooth}({\bf P})$)\;
 
\For{$t \gets 1$ \textbf{to} $T$}{
    \If{$\psi < P_t[S_t] <  \max_c P_t[c]$}{
        $\omega_t \gets \gamma$ {\bf if} $S_t$ is {\tt blank} {\bf else} $\omega {\bf p}$ [\text{word index of }$S_t$]\; 
        $P^s_t \gets (1-\omega_t)P_t + \omega_t\text{oneHot}( S_t)$\;
    }
    \Else{
        $P^s_t \gets P_t$\;
    }
}
$\vP^s \gets P^s_1,\ldots,P^s_T$\;
\Return{Beam-decode using $\vP^s$ and local LM of $\cC$}
\caption{The \algname\ Inference Algorithm}
\label{algo:mix_prob}
\end{algorithm}


%
\paragraph*{Aligning service characters}
Our first step is to expand out the characters in $\vs$ over the $T$ frames by repeating characters or inserting blanks so as to maximize the probability of the aligned characters as per $P_1,\ldots,P_T$.  Let $S$ denote the highest probability expanded character sequence.  
An example is shown in Table~\ref{tab:example} where $\vs={\tt posted}$ is aligned over $T=10$ frames and the resulting $S$ is shown in the first row. The full $\vP$ cannot be shown but we show the probability of the aligned character below it and the maximizing character probability in the second row. 
Such a forced alignment of $\vs$ with $P$ can be  solved optimally using a simple  Viterbi-like dynamic programming algorithm.  The algorithm processes $\vs$  time-synchronously over the $T$ frames such that either a symbol from $\vs$ or a blank is produced as output at each frame. This is referred to as ``Viterbi-align" in Algorithm~\ref{algo:mix_prob}.
Successfully aligning the service characters requires  an additional consideration. The server's output $\vs$ contains characters that can be attributed to both accent errors and cascaded language model errors. We therefore smooth $\vP$ distribution by adding a small constant $10^{-20}$ to all probability entries so even unheard characters get non-zero probability. 

\paragraph*{Revising $\vP$ with $\vs,\vp$}
Now each frame $t$ is aligned with a character $S_t$ in service.  We need to revise $\vP$ so as to 'support' the aligned characters of service but while ignoring those characters which may have been erroneously introduced during LM-based decoding.  For this we boost the probability of that character in $P_t$ on those frames $t$ where the probability $P_t(S_t)$ is less than the maximum probability in $P_t$ but greater than a threshold $\psi$.  
The lower limit $\psi$ is to suppress those characters in $\vs$ which are not 'heard' at all by the client's acoustic model, and are likely to have been introduced by the LM.  The amount of boosting is product of a hyper-parameter $\omega$ and the confidence of the parent word of $S_t$. If $S_t$ is blank, we use a fixed probability $\gamma$.  We use $\vP^s$ to denote the $\vP$ distribution after this revision with $\vs$.  
In Table~\ref{tab:example} we show the mode of the revised distribution $\vP^s$ in row 3.  Note, how 'p' in frame 2 was ignored in favor of the gold character 't' since $P_2(p)$ has a very small probability (1e-11).
In frame 8, the 'e' from the service was used to boost the probability of 'e' in the $P_8$ distribution from 0.44 to 0.66. Likewise in frame 9. Greedy decoding on the revised distribution yields {\tt to\,sted} which is closer to the gold token {\tt toasted} than either the service token {\tt posted} or the local token {\tt to\,state}.  Beam-decoding on the revised $\vP^s$ recovers the gold token. 


The above merging algorithm is simple and requires tuning only three hyper-parameters.  Since client's labeled data is limited, we found that more elaborate attention-based merging models using several parameters did not perform well. 
%




\section{Experiments}
We evaluate \algname\ on three accents and three service combinations and contrast against four other methods. We present anecdotes and analyze the kind of accent adaptations we achieve.\footnote{code available at https://github.com/Kartik14/FineMerge}
\paragraph*{Datasets}
We used the Mozilla Common Voice v4 (MCV-v4) dataset. The dataset is crowd-sourced and contains 1,118 hours of validated speech data of varying accents. We got around 28K Indian, 27K Australian and 63k British accented utterances, amounting to $37$, $35$ and $80$ hours of speech, respectively. For each accent, we split into train, validation and test sets roughly in the ratio 85-5-10 ensuring no overlap among speakers and transcripts. The MCV-v4 audio clips were normalized in a pre-processing step. 

\paragraph*{Service and Local Models}
We used Google Cloud Speech to Text API \cite{GoogleASR} as our default service model, and include two other service models later.
For the local, we used the DeepSpeech2 (DS2) \cite{DS2} model
pretrained on the LibriSpeech corpus \cite{Librispeech} and then fine-tuned individually for each accent. We used a trigram LM  trained on sentences from the MCV-v4 corpus after removing sentences overlapping with test sets. DS2 parameters $\alpha$ (for LM weight) and $\beta$ (to encourage more words) were also fine-tuned on the validation set for each accent.
The hyper parameters of our method $\omega$, $\psi$, $\gamma$ were also tuned on the validation set for each accent. 
%
\paragraph*{Methods compared}
We measure word error rates (WER) on five different models: the service model, the local model, Rover~\cite{rover} on the confidence weighted transcripts of service and local model, LM rescoring top-N whole transcripts from service, and our \algname\ method.  
\paragraph*{Overall Results}
\label{sec:mainresults}
In Table~\ref{tab:overall} we show the WERs on the Indian and Australian accents on these five methods.  Observe that overall the error rate of Service is lower than that of accent-adapted Local.  Rover's word-level merging provides significantly improved results than either of the two indicating that the two models exhibit complementary strengths.  LM rescoring does not improve results much, establishing that the local LM may not have much impact on the improved results. Our algorithm \algname\ provides the greatest gains in WER over all methods.  For Australian, we obtain a 28\% relative reduction in WER over either of the service and client models.
\begin{table}[h]
    \centering
    \begin{tabular}{|l|r|r|r|r|r|r|} \hline
        Method & \multicolumn{3}{|c|}{WER} & \multicolumn{3}{|c|}{CER} \\
        & Ind & Aus & uk & Ind & Aus & uk \\  
        \hline
        Local & 27.99 & 24.41 & 25.06 & 16.98 & 14.55 & 14
        28\\ \hline
        Service & 22.32  & 23.52 & 20.82 & 11.96 & 13.27 & 11.20 \\ \hline
       Rover & 21.12 & 18.04 & 18.10 & 11.95 & 9.81 & 9.88 \\ \hline
       LM rescore &  22.10 & 23.42 & 20.96 & 12.10 & 13.56 & 11.56\\ \hline
       \algname & {\bf 18.45}  & {\bf 16.90} & {\bf 16.47} & {\bf 10.65} & {\bf 9.33} & {\bf 9.79} \\ \hline
    \end{tabular}
    \caption{Overall comparison on WER and CER for Indian, Australian and British Accented Data}
    \label{tab:overall}
\end{table}
Table~\ref{tab:anec} presents some anecdotes which 
show how the fine-grained merging enables us to recover the highlighted word, even when neither the service nor client models contain that word.

\begin{table}[]
\setlength\tabcolsep{2.0pt}
    \centering
    \begin{tabular}{|l|l|l|} \hline
     & INDIAN & AUSTRALIAN \\ \hline
Gold & everyone toasted the ..  & nora finds herself ugly ..\\
Service & everyone posted the .. & nora van to self ugly ..\\
Local & everyone to state the .. & nor iphones herself ugly ..\\
Rover & everyone to posted the ..  & nor to self ugly ..\\
\algname & everyone {\bf toasted} the .. & nora {\bf finds} herself ugly .. \\ \hline 

Gold& for a brief time .. &  hannelore is an ..\\
Service& soda beef time .. &  i don't know what is an ..\\
Local & for a breese time .. &   hailar is an ..\\
Rover & for a beef time .. &  i don't know what is an ..\\
\algname & for a {\bf brief} time .. &  {\bf hannelore} is an ..\\ \hline

Gold& the condition also occurs.. & ..rope a bull while on a \\      
Service& definition of circus.. &  ..work a bowl while on a \\     
Local & the condition also acres.. &  ..rope the ball while on a \\     
Rover & the definition also circus.. &  ..work a bowl while on a \\     
\algname & the condition also {\bf occurs} & ..rope a {\bf bull} while on a \\ \hline  




\end{tabular}
\caption{Anecdotes comparing transcripts of Indian and Australian accents speech from five different methods.}
\label{tab:anec}
\end{table}


\paragraph*{Comparing methods of character alignment} A centerpiece of our method is  Viterbi aligning $\vs$ with the frame-level character probability distribution.  We show that this achieves a character-level alignment that is more accurate than existing methods by focusing only on character error rate (CER) before beam-decoding.  The last two columns in Table~\ref{tab:overall} presents CER of Local (before LM decoding), Service (as is), Rover applied at the character-level on these two, LM rescoring, and \algname's after selecting the modes of the revised distribution $\vP^s$ i.e., before LM decoding.  We observe that \algname's CER is much lower particularly for Indian accent.  This explains that the main reasons for our gains is due to our novel frame-level fine-grained merging algorithm.




\paragraph*{Varying Quality of Service Model}
In addition to the default Google Speech API service (G-US), we evaluate on two other models as service --- a second Google speech-to-text model (G-Video) \cite{GoogleASRVideo} meant for transcribing audio of video files, which works significantly better for MCV-v4 utterances because of their low-fidelity, and Jasper~\cite{li2019jasper}, a recent end-to-end convolutional neural ASR model trained on the LibriSpeech dataset. %
We note here that we opted for G-US rather than Google's ASR API for Indian English because of the latter's poor performance (compared to G-US) on MCV-v4 utterances that are low bandwidth.
Table~\ref{tab:service} shows the results.  WER of Local stays the same since service has no role during its training. We see a wide difference in accuracies across the different services.  G-Video is the most accurate, but even in this case \algname\ is able to obtain a relative WER reduction by at least 3\%.  The Jasper model is worse than local Indian fine-tuned, yet \algname\ achieves more than 15\% relative WER reduction wrt both service and local.  This shows that the hyper-parameters of our service guided local inference adapt even to a weaker service.
\begin{table}[h!]
\setlength\tabcolsep{3.0pt}
    \centering
    \begin{tabular}{|l|r|r|r|r|r|r|} \hline
        Method & \multicolumn{3}{|c|}{Indian} & \multicolumn{3}{|c|}{Australian} \\
        & G-US & G-Video & Jasper & G-US & G-Video & Jasper \\
        \hline
        Local &  \multicolumn{3}{c|}{~~~~27.99} & \multicolumn{3}{c|}{~~~~24.41}   \\ \hline
        Service & 22.32 & 13.77 & 31.82  & 23.52 & 11.08 &  19.56 \\ 
        \hline
       Rover &  21.12 & 20.51 & 26.95 & 18.04 & 13.84 & 17.57  \\ \hline
      LM rescore & 22.10  & 13.37 & 31.38 & 23.42 & 10.99 & 19.35 \\ \hline
       \algname & {\bf 18.45}  & \textbf{13.36} & {\bf 23.72} & {\bf 16.90} & {\bf 10.68} &  {\bf 16.07} \\ \hline
    \end{tabular}
    \caption{Effect of changing service model}
    \label{tab:service}
\end{table}

\paragraph*{Importance of Accent Adaptation}
One interesting question was if our gains were merely due to ensembling of any two independent models adapted to test data domain, or did we specifically adapt accent.  To answer this, we run \algname\ with a local model fine-tuned on a similarly-sized MCV corpus from a different accent.
Table \ref{tab:Importance of accent} compared our WER to  the WER obtained after the client models for the two accents is finetuned on US accented sample.

Observe that \algname\ out-performs service even when the local model is fine-tuned on a different accent.  This captures the base benefit of ensembling.  However, after fine-tuning on data of its own accent the gains are higher.  For Aus accent, service WER of 23.52 drops to 20.66 with \algname\ on  Indian-local but drops further to 16.90 on Aus-local.
\begin{table}[]
    \centering
    \begin{tabular}{|l|c|c|c|}
        \hline
        Test & Service & \multicolumn{2}{|c|}{\algname\ with }  \\ 
           accent  &             & (ind/aus)-Local & us-local
           \\ \hline
        Indian &  22.32 & 18.45 & 21.01\\
        Aus & 23.52 & 16.90 & 20.66\\ 
        \hline
    \end{tabular}
    \caption{WER comparison with different local models.}
    \label{tab:Importance of accent}
\end{table}



\begin{figure}[t!]
    \centering
    \includegraphics[width=0.8\linewidth,height=0.65\linewidth]{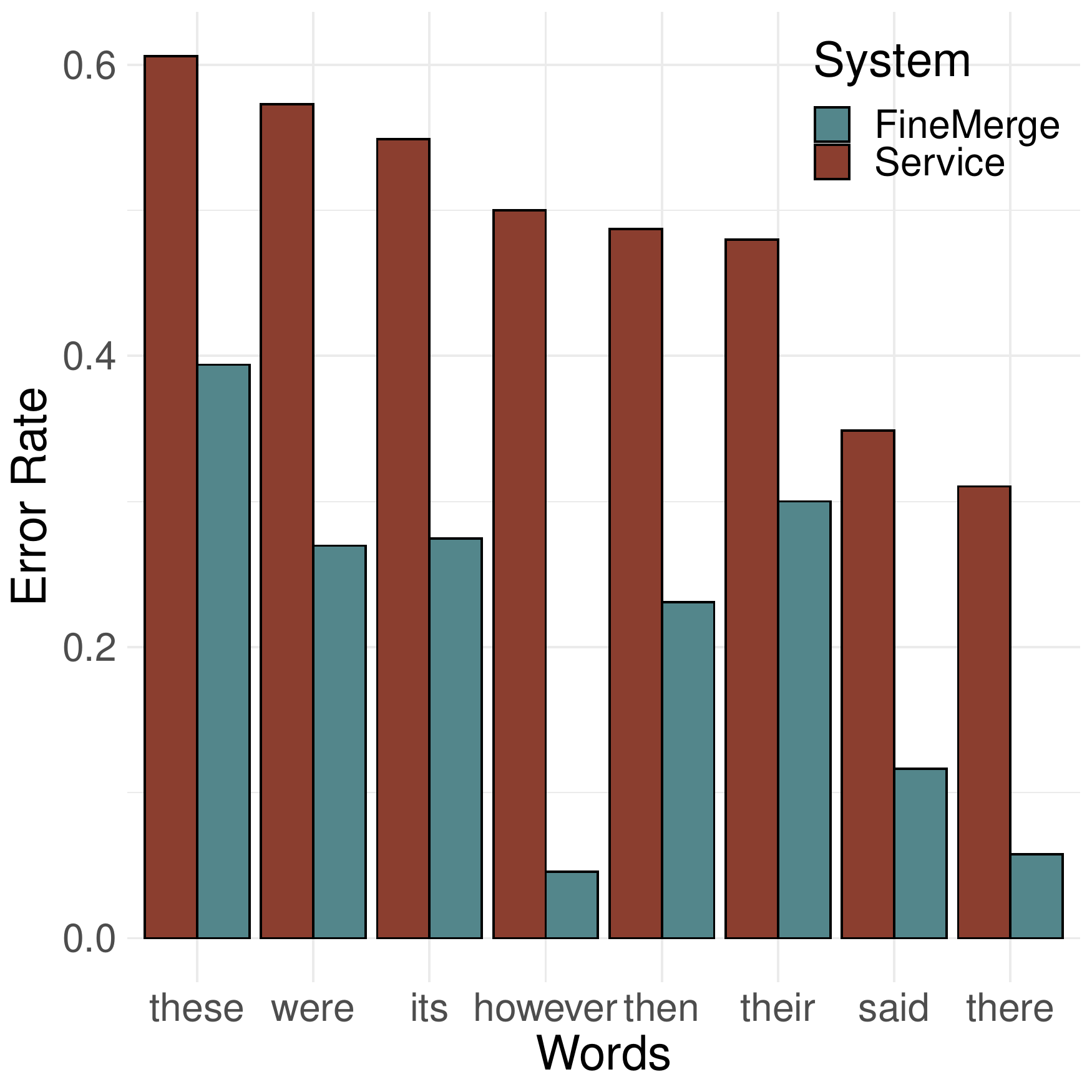}
    \caption{Highest reductions in error per word  on Indian-accented test samples.}
    \label{fig:freqs}
\end{figure}


Figure~\ref{fig:freqs} shows the largest reductions in errors per word on Indian test samples obtained by \algname\ over service.
%
Error rates are cut in half for most words
revealing \algname's ability to do accent adaptation. Word 
``however" is an interesting example to highlight. The diphthong /AW/ in ``however" has a wide range of phonetic realizations across Indian speakers; 
and has been investigated in prior work~\cite{maxwell2010acoustic}. This variability is difficult for the service to accurately model, while \algname\ cuts the errors on ``however" down to $5\%$ from $50\%$. Another interesting
example is ``were". The phonemes /v/ and /w/ are indistinguishable in most Indian languages, making minimal pairs like \emph{veil} and \emph{wail} homophones when articulated by Indian speakers. 
/DH/-initial words like ``then", ``these", ``their" and ``there" are other likely targets of accent errors due to the lack of dental fricatives like /DH/ in most Indian languages. \algname\ is able to substantially reduce these errors.

\section{Conclusion and Future Work}
In this paper we motivated and introduced the problem of black-box adaptation of an ASR service. We presented a novel coupling of an open-source accent adapted model with the black-box service model to fix accent errors in an otherwise strong service model.  We presented \algname\ an algorithm that achieves a fine-grained mixing of the service output and local frame-level distributions.  We show that such fine-grained mixing is specifically effective in fixing accent errors that word-level mixing cannot fix.  Our strategy achieves upto 28\% reduction in word-error rate over service APIs of varying grades of quality. Future work could consider combining outputs from multiple services and fixing both dialect and accent differences.

\newpage

\bibliographystyle{IEEEtran}
\bibliography{references}


\end{document}